\documentclass[apj]{emulateapj}
\usepackage{epsfig}
\usepackage{multirow}
\usepackage{graphicx}
\usepackage{epstopdf}
\usepackage{amsmath, amsthm, amssymb}
\usepackage{natbib}
\usepackage{rotating}
\shorttitle{Coupled stellar motions in the Galactic Disk}
\shortauthors{D'Onghia et al.}
\graphicspath{{figs/}}

\begin{document}

\title{Excitation of coupled stellar motions in the Galactic Disk by orbiting satellites}

\author{E. D'Onghia\altaffilmark{1,2}, P. Madau\altaffilmark{3}, C. Vera-Ciro\altaffilmark{1}, A. Quillen\altaffilmark{4}, L. Hernquist\altaffilmark{5}}

\begin{abstract}
We use a set of high-resolution N-body simulations of the Galactic
disk to study its interactions with the population of satellites
predicted cosmologically. One simulation illustrates that multiple
passages of massive satellites with different velocities through the
disk generate a wobble, having the appearance of rings in face-on
projections of the stellar disk. They also produce flares in the disk
outer parts and gradually heat the disk through bending waves. A
different numerical experiment shows that an individual satellite as
massive as the Sagittarius dwarf galaxy passing through the disk will
drive coupled horizontal and vertical oscillations of stars in
underdense regions, with small associated heating.  This
experiment shows that vertical excursions of stars in these
low-density regions can exceed 1 kpc in the Solar neighborhood,
resembling the coherent vertical oscillations recently detected
locally.  They can also induce non-zero vertical streaming motions
as large as 10-20 km s$^{-1}$, consistent with recent observations in
the Galactic disk. This phenomenon appears as a local ring, with modest
associated disk heating.


\end{abstract}


\altaffiltext{1}{Department of Astronomy, University of Wisconsin, 2535
  Sterling Hall, 475 N. Charter Street, Madison, WI 53076, USA.
  {e-mail:edonghia@astro.wisc.edu}}

\altaffiltext{2}{Alfred P. Sloan Fellow}
\altaffiltext{3}{University of California Santa Cruz, 1156 High Street, Santa Cruz, CA 95064}
\altaffiltext{4}{Department of Physics and Astronomy, University of Rochester 
Rochester, NY 14627}
\altaffiltext{5}{Harvard-Smithsonian Center for Astrophysics 60 Garden St., MS-51}

\section{Introduction}
\label{sec:intro}

The nature of the vertical bulk motion of stars recently discovered in
the solar neighborhood remains uncertain. These motions assume the
form of vertical oscillations of the stellar disk as found by the
RAdial Velocity Experiment (RAVE) and in the Large Area Multi-Object
Spectroscopic Telescope (LAMOST) radial velocity survey
\citep{Widrow12,Williams13,Carlin13}.  In addition, \citet{Widrow12}
and \citet{Yanny13} find evidence for wave-like North-South
asymmetries in the number counts of stars in the solar neighborhood.
The origin of these non-zero average vertical velocities is
controversial and different studies advocate for a variety of
formation scenarios, from being caused by the transient nature of
spiral structure \citep{Debatt14}, to the presence of the bar
\citep{Monari14,Monari15}, to external perturbations, such as
the Sagittarius dwarf galaxy \citep{Widrow14,Gomez12,Gomez13,Gomez16,Vega15}.


In a hierarchical universe, in which disks are believed to form and
evolve, galaxies are surrounded by a population of orbiting luminous
and dark satellites, with $\sim 10-15$\% of them expected to pass
through the primary disks \citep{Don10a} and perturb them.  In
particular, the interaction of a stellar disk with visible dwarf
satellite galaxies has been suggested as a way to excite structures in
galactic disks, such as rings, warps, spiral structures, as well
providing mechanisms for dynamically heating stellar disks
\citep[][]{Quinn93,Walk96,Chakra09,Young08,Struck11,Stelios09,Purcell11,Bird12,Gomez12}.

An early study by \citet{TH92} computed the energy deposited locally
into a disk and halo through a merger with a satellite galaxy on a
nearly circular orbit, decaying by dynamical friction.  However,
self-gravitating disks support collective modes, such as density waves
propagating across and perpendicular to the plane, which are neglected
in the impulse approximation or dynamical friction formalism, but are
naturally described in N-body experiments.  Later numerical
simulations indicated that disks could be strongly heated by the
accretion of massive satellites and included other effects, previously
ignored, such as the tidal disruption of the satellite as it
approaches the host galaxy and the tilting of the disk plane due to
the gravitational torque from the satellite.  However, some of the
early calculations suffered from numerical relaxation and noise
\citep{Sell13}.  Low mass and spatial resolution make a disk unable to
support collective modes, casting doubts on the outcome of the
numerical experiments.

\begin{figure*}[th]
  \begin{center}
    \includegraphics[width=0.9\textwidth]{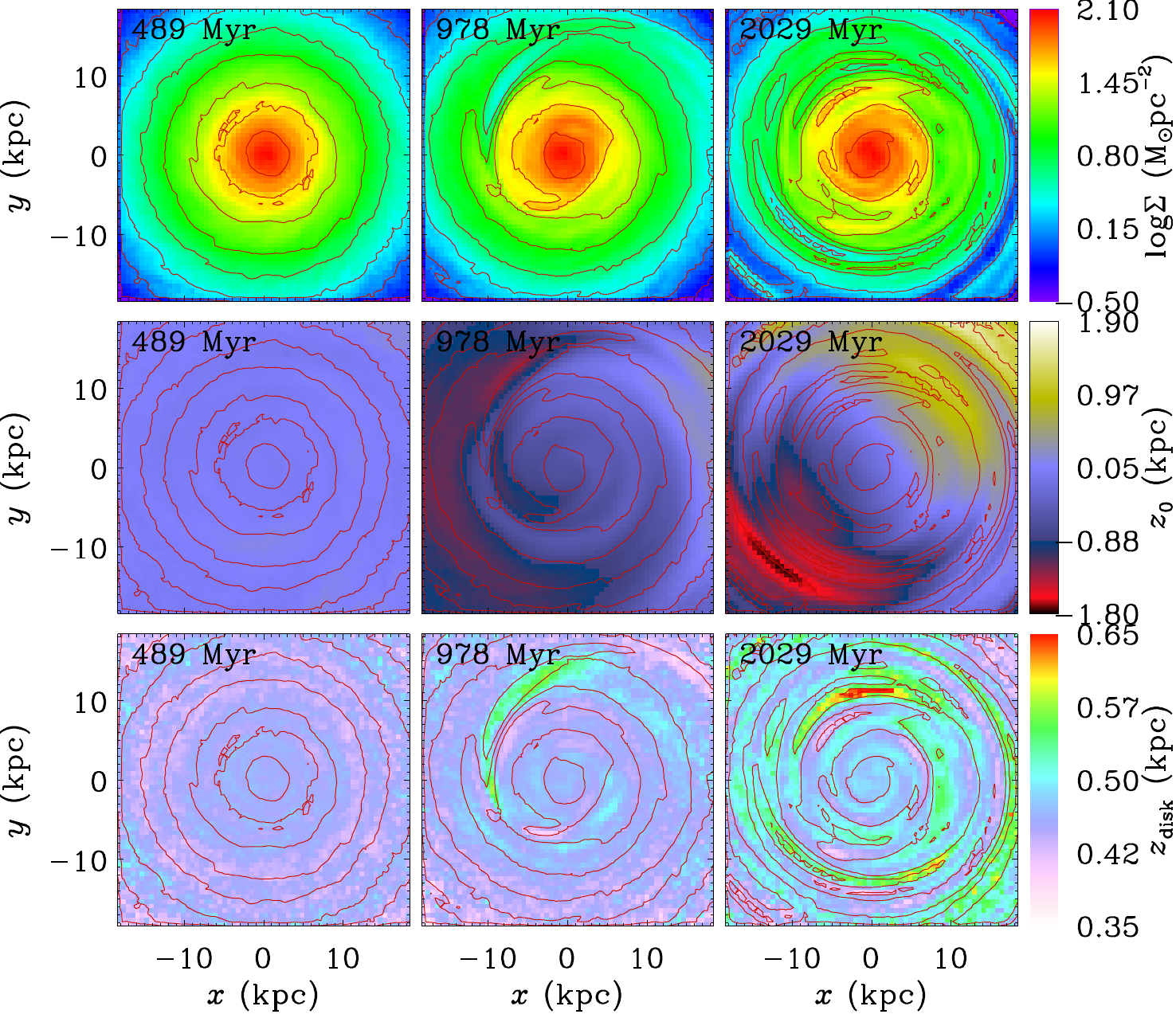}
  \end{center}
  \caption{Time evolution of the stellar disk seen face-on
    perturbed by one satellite of mass
    2x$10^{10}$ M$_{\odot}$. The surface mass density values are displayed (top panels).
    Time evolution of the same disk face-on with the first moment of the vertical 
    component of the stars shown (middle panels) and 
    enhancements of the thickness of the disk 
    displayed with their values (bottom panels).}
\label{fig:evolution}
\end{figure*}

The vertical structure and thickening of galaxy disks due to accretion
of small satellites was explored by \citet{Sell98}, who suggested that
this process can excite large-scale bending waves in a disk.
Associated vertical heating involves dissipation of bending waves at
resonances with stellar orbits in the disk.  More recent numerical
experiments found that under realistic circumstances the vertical
heating of disks scales with the satellite to disk mass ratio as $\sim
(M_{sat}/M_{disk})^2$ \citep{Hopk08}.  Therefore, it is expected that
disk heating is caused primarily by the high-mass end of the satellite
distribution predicted in $\Lambda$CDM cosmological simulations and is
reduced in the solar neighborhood \citep{Just15} relative to early
estimates \citep{Hopk09}.


One assumption often made in studies of disk dynamics is that vertical
heating of a disk is largely unaffected by the energy deposited into
motion in the plane and that vertical and horizontal motions are not
coupled.  Based on this assumption, the vertical heating of stellar
disks has always been considered a relatively simple problem, since
the stars are not expected to be influenced in their vertical
oscillations by the energy deposited into their radial motions.

In this paper we study the effects of the passage of satellites
through stellar disks when self-gravity is included, and examine
consequences for disk-thickening and kinematic signatures for stars in
the solar neighborhood.  We show that when at least one massive
satellite impacts the disk, the horizontal and vertical motions
locally {\it are} coupled, having implications for the vertical
structure and the heating of stellar disks.

\section{NUMERICAL METHODS}
\label{sec:methods}

\begin{figure*}[th]
  \begin{center}
    \includegraphics[width=0.9\textwidth]{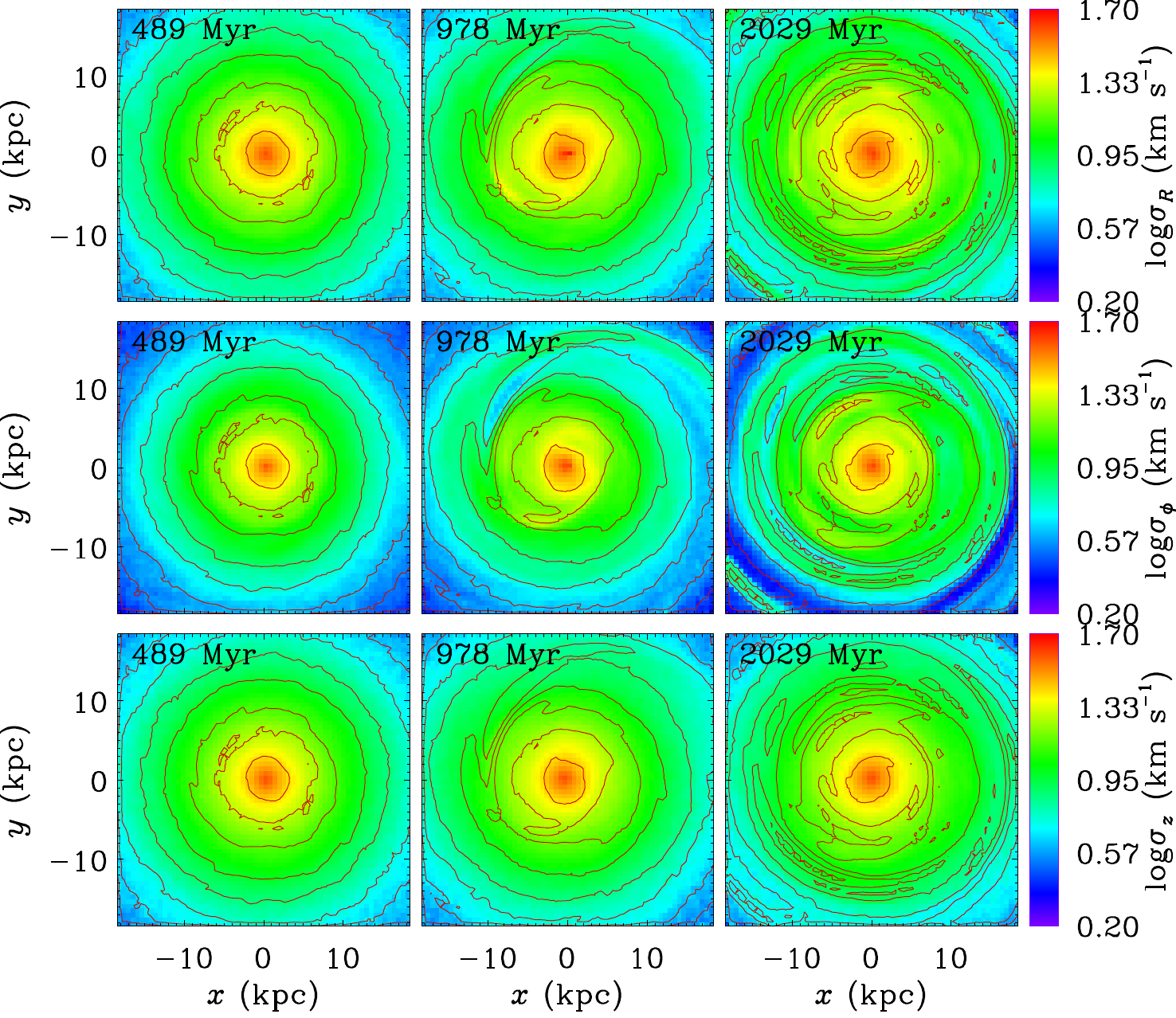}
  \end{center}
  \caption{Time sequence of the stellar velocity dispersion: radial (top panels), tangential
  (middle panels), and vertical (lower panels) of the stellar disk as
    perturbed by a satellite as massive as 2$\cdot 10^{10}$ M$_{\odot}$ (first
    numerical experiment). There is a negligible increase
    of random motions in the in-plane motion of stars or
vertical heating of the disk, even after more than 2.5 Gyrs.}
\label{fig:face}
\end{figure*}

To investigate the gravitational response of a disk to the impact of
dark satellites, we have performed in total three N-body experiments
in which a live disk of 10$^8$ star particles was embedded in a 
spherical dark
halo with the introduction of one satellite dark galaxy as massive as
Sagittarius and in subsequent experiments with many substructures as
extracted from cosmological simulations.

\begin{table}[htbp]
\caption{GALAXY STRUCTURAL PARAMETERS} 
   \begin{tabular}{@{} lccr @{}} 
      \hline
      Parameters    & Experiment 1 &  Experiment 2 & Experiment 3  \\
      \hline

      a [kpc]$^{a}$     & 29.6  & 29.6  & 29.6 \\
      R$_d$ [kpc]$^{b}$ & 4.0   & 3.13  & 3.13 \\
      M$_{DM}$[M$_{\odot}$]$^{c}$ & 9.5x10$^{10}$ & 9.5x10$^{10}$  & 9.5x10$^{10}$ \\
      M$_{disk}$ [M$_{\odot}$]$^{d}$ & 1.8x10$^{10}$ & 1.8x10$^{10}$ & 1.8x10$^{10}$ \\
      z$_{disk}$ [kpc]$^{e}$  & 0.4 & 0.313 & 0.313 \\
      N$_{sat}$ $^{f}$   & 1   & 1,000 & $\sim$100\\
      M$_{sat}$ [M$_{\odot}$]$^{g}$ & 2x10$^{10}$ & 9.5x10$^{6}$ & (*) \\     
      \hline
\end{tabular}
\noindent
$^{a}$Halo scale length. $^{b}$ Stellar disk length. 
$^{c}$ Dark matter halo total mass. $^{d}$ Stellar disk total mass. 
$^{e}$ Disk scale heigth. $^{f}$ Number of satellites. $^{g}$ Satellite mass. 
$^{(*)}$ The satellite mass distribution 
follows a power law: dN/dM $\propto$ M$^{-1.9}$. $
\ \ \ \ \ \ \ \ \ \ \ \ \ \ \ $
   \label{tab:ic_params}
\end{table}

The dark-matter distribution of the host galaxy is modeled with a
Hernquist profile \citep{Her90} with a total halo mass and a
scale-length of mass 9.5x$10^{11}$ and 29.6 kpc, respectively (see Table 1).

The stellar disk in the initial conditions of all experiments is
described by an exponential surface density profile:
$\Sigma_{*}$(R)=$M_{*}$/(2$\pi$R$^2_d$) exp$^{(-R/R_d)}$, with R$_d$ the 
scale-length.  The vertical mass distribution of the stars
in the disk is specified by the profile of an isothermal sheet with a
radially constant scale height $z_{disk}$.  The 3D stellar density in
the disk is then:


\begin{eqnarray}
 \rho_{*} (R,z) &=& \Sigma(R) \zeta(z) \nonumber \\
 &=& \frac{M_{*}}{4 \pi \rm{z_{disk}} \rm{R_d^2} } \rm{sech}^2
\Big(\frac{z-z_0}{z_{disk}}\Big) \rm{exp} \Big(-\frac{R}{R_d}\Big)
\end{eqnarray}

\noindent
Here, $z_0$ is the first moment of the $z$-component of stars,
$z_{disk}$ is treated as a free parameter that is set by the vertical
velocity dispersion of the stars in the disk, and the velocity
distribution of the stars is chosen such that this scale height is
self-consistently maintained in the full 3D potential of the galaxy
\citep{H93,SdMH05}.  A value of $z_{disk}=0.1$ R$_d$ is adopted.  The
parameters describing the total halo mass and the stellar disk are
summarized in Table 1.  The details of
the rotation curve of each galaxy component and the behavior of the
Toomre parameter $Q$ as a function of radius is reported in Fig.1 of
\citet{Don13} and \citet{VC14}. The disk mass fraction within two 
scale-lengths is 30\%. With this choice for the structural parameters
the outcome is a multi-armed spiral disk galaxy with no stellar bar \citep{D15}.

Unlike in \citet{Don13}, where the disk was perturbed by co-rotating
overdense regions with masses similar to giant molecular clouds, here
instead we have explored the dynamical response of the disk to
interactions with one or many substructures.

{\it First Experiment}. We added to the stellar disk only one live substructure 
as massive as the Sagittarius dwarf galaxy, with a mass of 2x$10^{10}$ M$_{\odot}$
and hitting the disk at roughly at 2.2 kpc from the disk center. 
The disk was surrounded by a live dark-matter halo in this case.  The satellite was placed
on an eccentric orbit (e=0.7) at an initial distance of 20 kpc below
the plane. It completed one orbit around the disk in approximately 1.7
Gyrs.
 
{\it Second Experiment}. We added 1,000 substructures with a mass of 
10$^{7}$ M$_{\odot}$, hitting the disk from random orbits, to the live disk embedded in a static potential for the halo.

{\it Third Experiment}. In this experiment the same galaxy model was evolved but with the introduction of $\approx$ 100 live dark satellites
isotropically distributed. In this set-up we have at least two satellites with mass larger
than 10$^{9}$ and several satellites with mass larger than 2x10$^8$
M$_{\odot}$ that pass through the disk.

The $\approx$ 100 satellite halos have masses between M$_{sat} = 2
\rm{x} 10^8$ −- 2x10$^{10}$ M$_{\odot}$ orbiting around a stellar disk
with a mass of 1.9x$10^{10}$ M$_{\odot}$ embedded in a static dark
halo.  The mass distribution of the substructures is characterized by
a power law d$N/$d$M \propto M^{-1.9}$ as expected from cosmological
simulations \citep[][]{Diemand08,Springel08}. The cumulative mass of
all subhalos is 10\% of the total mass of the galaxy halo.

Each substructure consists of a live dark-matter halo modeled assuming
a Hernquist profile, with a scale-radius, $a$, related to the NFW
profile through the relation \citep{SdMH05}:
$a=r_s\sqrt{2[ln(1+c)-c/(1+c)]}$, where $c$ is the subhalo
concentration as predicted in cosmological simulations.  The number of
particles ranges from 10$^3$ to a maximum of 5x10$^5$ particles for
the most massive substructure.


\noindent

\section{Results}
\label{sec:results}

\subsection{Dynamical Response of the Disk to Satellite Perturbations}

We have verified first that low-mass disks run with sufficiently large
numbers of particles do not develop spiral structures in the absence
of any perturbing influence \citep{Don13}.

\begin{figure}[th]
  \begin{center}
    \includegraphics[width=0.5\textwidth]{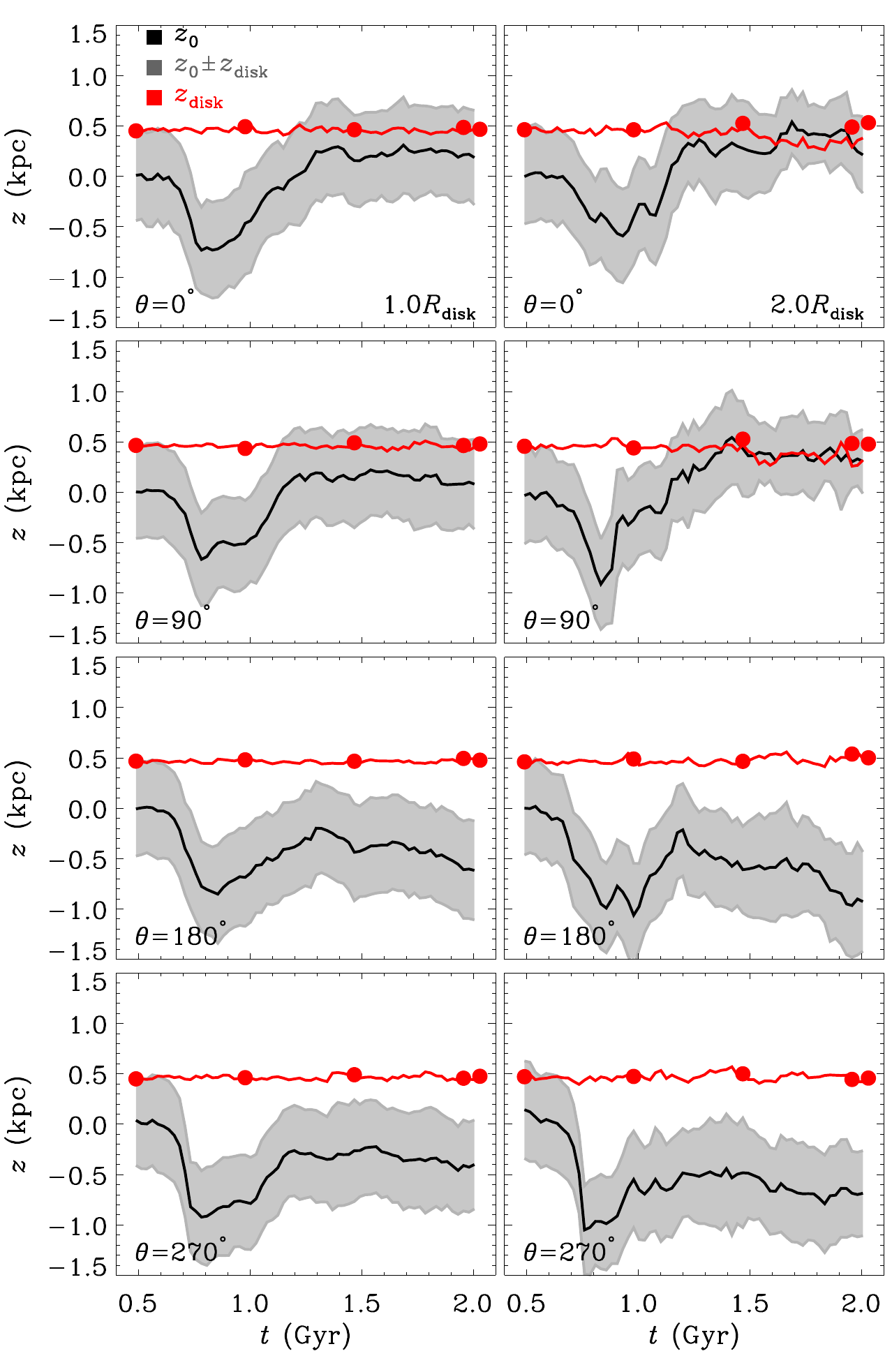}
  \end{center}
    \caption{{\it Left Panels}. Time sequence of the average vertical component
    (solid black line) of stars located at  R$_{d}$=4 kpc, for different azimuthal angles:
   $\theta=0;90;180;270$ for the stellar disk perturbed by one massive
      satellite with M=$2\cdot 10^{10}$ M$_{\odot}$ (described as first
      numerical experiment). 
    The gray area is the 1-$\sigma$ vertical
    displacements around the average value. The disk thickness is also displayed as a
    function of time (solid red line). {\it Right Panels}. The same sequence
    as above is displayed for stars located at 2R$_{d}$=8 kpc.}
\label{fig:face}
\end{figure}

Fig.1 shows the outcome of the first experiment where the disk is
displayed face-on after being hit by a massive dark-matter
substructure.  We note that the disk initially develops an open,
two-armed pattern that propagates outwards to form a ring. The
satellite is massive enough that its impact causes the disk to respond
immediately after the intruder passes, with the stars around the
collision site moving inwards and then expanding outwards.  As a
result of the contraction and expansion there is a wave of enhanced
density which moves and forms a pattern similar to a ring (top panels
of Fig.1).  Indeed, this process is thought to be responsible for
rings in galaxies through off-center collisions, but not the
structures that resemble the patterns we see in nearby spiral galaxies
\citep[][]{Lynds76,HW93,DMM08,MM12}.  Similar features are produced in
the simulation following several satellites passing through the disk.

To better understand the forming pattern in our simulations we
consider a heuristic model. We assume a displacement of the disk in
$z$ with a function proportional to e$^{im\phi}$. Then the ring
pattern rotates with a pattern speed:

\begin{equation}
\Omega_p = \Omega - \frac{\nu_z}{m}
\end{equation}

\noindent
where $\nu_z$ is the halo vertical oscillation frequency (see
the derivation below for the Hernquist model), and $\Omega(R)$ 
is the angular rotation rate. 
For $m = 1$ we expect $\nu_z(R) > \Omega(R)$
with a {\it retrograde} ring pattern. 

Note that the pattern is expected to move with angular rotation rate ($\Omega - \nu_z$), 
with its derivative with respect to radius being the sheer rate for the vertical 
disturbance. For an initial 
perturbation at a given azimuthal angle, with size $\Delta R$, 
after a time $\Delta t$ the perturbation  will be wrapped in $2 \pi$ across a radial distance: 

\begin{equation}
\Delta t \sim \frac{2 \pi}{(\nu_z - \Omega) \Delta R} \approx R \frac{T}{\Delta T}
\end{equation}

\noindent
where T is the orbital period for stars.  At R=2.2 kpc, T=0.12 Gyr. 
The estimate of the time for a {\it ripple} to appear at 2.2 kpc from the galaxy
center is $\sim 200$ Myrs as indicated in our simulations giving $\Delta R=1.32$ kpc. 

Fig.1 illustrates, in its top panels, the time evolution of the disk
displayed face-on for approximately 2 Gyrs.  Next we computed the the
first moment of the $z$ component of stars: $z_0=\Sigma z_i/N$, with N
being the number of stars at that radius and from this the
characteristic scale height of the disk, $z_{disk}$ during its
evolution.  To do this, it is convenient to parametrize the function
$\zeta (z)$ of eq.(1) such that $\int dz \zeta (z)=1$ \citep{BT08}.
With this choice, the quantity $\Sigma (R) \zeta(z) dz$ is the surface
density of the layer of stars with thickness $dz$ that lies at
distance $z$ from the midplane.  Fig.1 shows a time sequence of the
face-on disk with a map of the the average vertical component, $z_0$
and the characteristic disk scale height, $z_{disk}$ (middle and
bottom panels, respectively).

\subsection{Vertical and horizontal stellar motions coupled}

As the disk wobbles and waves of enhanced density move and form
patterns similar to rings, we notice that the
underdense regions coincide with a local increase of the characteristic scale
height, $z_{disk}$, at that specific location in the disk plane as illustrated
in Fig.1.
\begin{figure*}
  \begin{center}
    \includegraphics[width=0.9\textwidth]{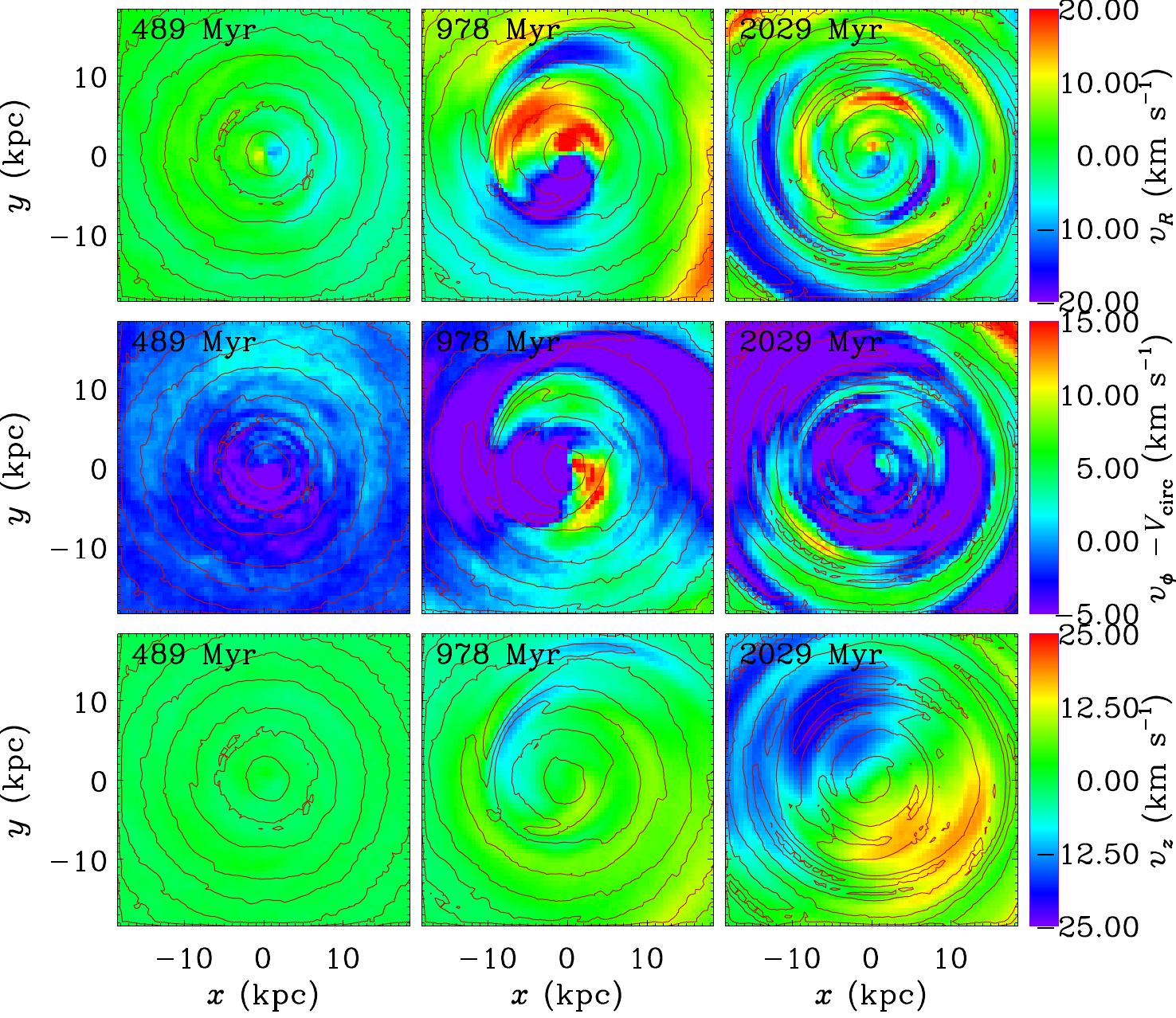}
  \end{center}
  \caption{Time evolution of the streaming velocity for the perturbed
    disk: radial (top panels), tangential (middle panels), vertical (bottom
    panels) computed for the case when the disk is perturbed by the passage of
    one satellite as massive as 2$\cdot 10^{10}$M$_{\odot}$ (first numerical experiment). 
There is a {\bf small} increase of random motions in the in-plane motion of stars
or vertical heating the disk, after more than 2.5 Gyrs.} 

\end{figure*}

In galactic disks, stars are always assumed to be on nearly circular
orbits, with the motion usually well-described by the epicyclic
approximation, and where the in-plane and vertical displacements of a
star are decoupled.  Stars near the disk midplane
oscillate vertically with a frequency $\nu_z^2=\frac{\partial^2
  \Phi}{\partial R^2}$ where $\Phi$ is the gravitational
potential. For a Hernquist density profile, the potential is:\\

\begin{equation}
\Phi(r)=-\frac{GM}{r+a}
\end{equation}
\noindent
where $a$ is the halo scale-length, $G$ is the gravitational constant and $M$ is the total halo mass.
Then the vertical oscillation frequency is:

\begin{eqnarray}
\nu_z^2 &=& \frac{\partial^2 \Phi}{\partial z^2} = GM \ \ \cdot \nonumber \\
  & &  \cdot \frac{a(x^2+y^2)+(x^2+y^2-2z^2)\sqrt{x^2+y^2+z^2}}{(x^2+y^2+z^2)^{3/2}(\sqrt{x^2+y^2+z^2}+a)^3}
\end{eqnarray}

\noindent
Note that the vertical frequency is usually greater than the epicyclic
frequency. This means that the stars oscillate in their radial motion
more slowly than in their vertical oscillations. This condition is
particularly valid for a star located in regions of enhanced density,
like in a density wave where, due to the stronger gravitational field,
it will be forced to oscillate fast relative to the disk midplane.
However, a star located in an underdense region feels a weaker
gravitational restoring force, and has its vertical frequency reduced
to roughly match the circular frequency, specifically: $\Omega \sim
\nu_z$.  Thus for stars preferentially located in underdense regions
the horizontal motion will be coupled in frequency to the vertical
motion.  The overall effect of this unexpected coupling is to produce
local enhancements of the characteristic disk scale height as
illustrated in Fig.1.


\subsection{Stellar vertical oscillations and heating}

Ideally, one would like to use the impulse approximation to study the
energy deposited into the disk by a satellite passage.  For a
satellite of mass $M_{sat}$ moving at relative speed $V_0$ and impact
parameter $b$, the mean change in the vertical kinetic energy of a
disk star during a single satellite passage with random orientation is
\citep[][]{Spitzer,Sell08,Don10b}:

\begin{equation}
\Delta E_z \propto z_{disk}^2 \Big(\frac{GM_{sat}}{b^3 \nu_z}\Big) \beta^2
\end{equation}
\noindent
with the parameter $\beta=2 \nu_z b/V_0$ being the characteristic passage time
of the satellite divided by the stellar orbital period.  
For a high velocity satellite the net vertical energy transferred is low, since   
it is nearly the same as for a free particle, and thus varies as $V_0^{−2}$. 
At low velocities, however,
the energy transfer is exponentially small.  
But, a self-gravitating disk supports collective modes, which
include vertical waves and horizontal density waves that are not
included in the impulse approximation.  When these collective effects
are properly accounted for (e.g. using numerical simulations of
sufficient resolution of self--gravitating disks with low noise), the
deposition of vertical energy into the stars in the disk by the
passage of a satellite has been demonstrated to be greatly reduced
relative to the impulse approximation \citep[][]{Sell08,Hopk08,Just15}.
\begin{figure}
  \begin{center}
    \includegraphics[width=0.4\textwidth]{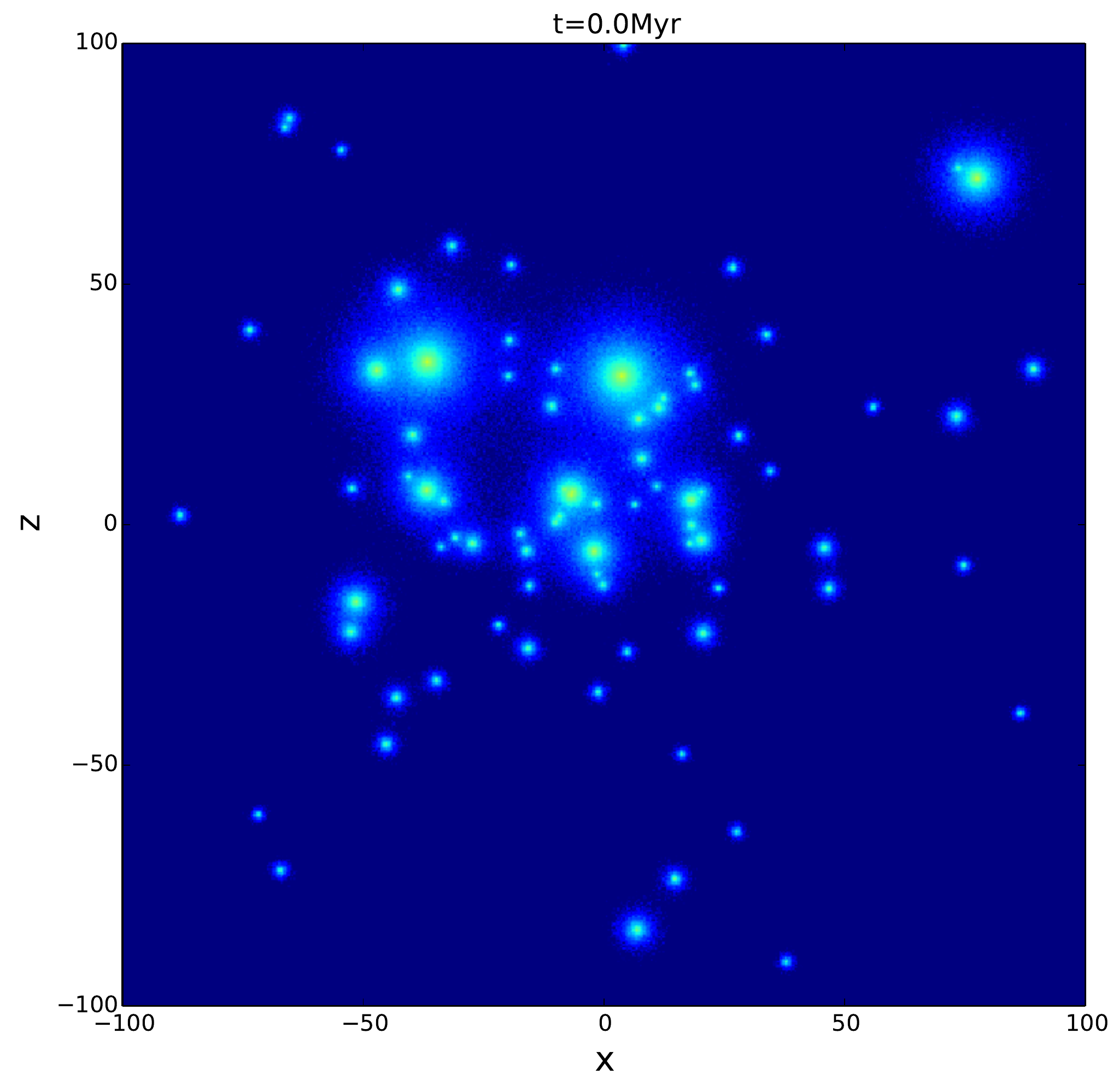}
  \end{center}
  \caption{The spatial distribution of $\approx 100$ live dark subhalos with
    structural properties inferred from cosmological simulations from 2x10$^8$
    M$_{\odot}$ to 2x10$^{10}$ M$_{\odot}$ (third numerical experiment).}
\label{fig:sub}
\end{figure} 

It has been argued that although a satellite passage does not heat a
thin disk, there is substantial heating at late stages, through the
process of exciting disk-bending waves. Once excited, these waves
eventually dampen efficiently at wave-particle resonances, thereby
heating the outer part of the disk \citep{Sell08}.  It should be noted
that this assumes that the in-plain motion is decoupled from the
vertical motion.

The outcome of our experiment with only the most massive satellite with 
M=2x$10^{10}$ M$_{\odot}$ hitting the disk is illustrated in Fig. 2, which displays 
a map of the radial, tangential and vertical velocity dispersion of stars 
at the time before  the satellite impacts the disk, and at later times.
It shows that there is some increase of random motions in the in-plane
  motion of stars or vertical heating of the disk, after more than 2.5 Gyrs.

To demonstrate our findings we evaluate the first moment of the z component of stars:
$z_0=\Sigma z_i/N$ and the second moment
$z_{disk}=\sqrt{\Sigma z_i^2/N}$, with $N$ being the number of stars at that radius.
Fig.3 displays the evolution in time of the first moment of the vertical component of the stars,
$z_0$, (solid black line) evaluated at each panel for different  azimuthal angles
($\theta=0,90,180$) and at different radii, $R=4$kpc (left panels) and $R=8$ kpc
(right panels), respectively.

It seems clear from our results that the satellite impact induces the disk to
wobble, with local enhancements of the characteristic scale length of the disk
in underdense regions, but with some increase of the disk thickness with time
as illustrated by the evolution of $z_{disk}$ in Fig.3 (solid red line).


Could features such as vertical displacements identified in our
simulations be a manifestation of bending waves or breathing modes?
Our N-body simulations involve self-gravitating disks; thus they
support collective modes. 
To address this question we focus on this run when only the most massive
substructure hits the disk, to
isolate the phenomenon from the cumulative effects of many encounters
with smaller substructures. If a massive satellite impacting
the disk excites the formation of bending waves, there are good
arguments to believe that these waves are characterized by long
wavelengths that dampen and produce heating in the outer parts of the disk \citep{Wein91}.

Instead, we notice vertical
displacements in all underdense regions that we attribute to the
effect of reduced restoring forces locally in the disk. This explanation is supported
by the fact that these enhancements in vertical motions involve a modest number of stars 
and with modest associated heating as shown in Fig.2-3.   


An interesting feature of our disk when hit by one massive satellite is that, at a distance
corresponding to the solar neighborhood, the vertical displacement of
stars in their motion perpendicular to the plane can extend up to 1
kpc from the midplane with asymmetries in the number density of stars above
and below the disk.  These features are confirmed by Fig.4
where maps of the velocity streaming motions--radial, tangential, and
vertical--for stars in the disk are given.  Note that Fig.4 shows that the
vertical streaming motions can be as large as 15-20 km$^{-1}$ and are
related to the wobbles in the disk.

\subsection{Dynamical response of the disk to many satellites}

\begin{figure}
  \begin{center}
    \includegraphics[width=0.4\textwidth]{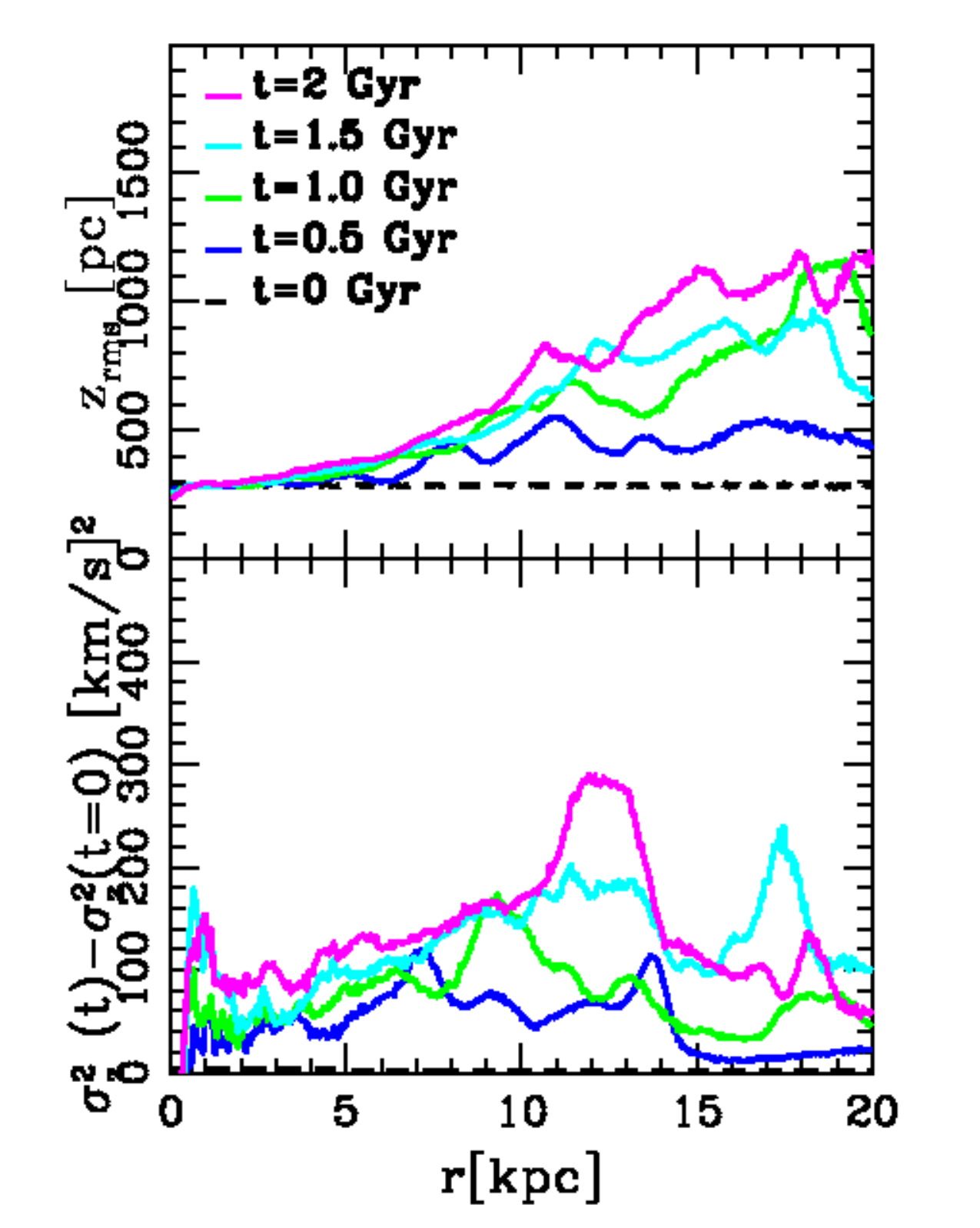}
  \end{center}
  \caption{{\it Top Panel}: Time sequence of the first moment of the z-component of stars as a function of radius
for the experiment with several satellites hitting the disk (third numerical experiment).
{\it Bottom Panel}: The change of the vertical random energy from the initial time.}
\end{figure}

We performed two additional 
N-body experiments in which a live disk of 10$^8$ star particles was
embedded in a rigid potential with the introduction
of substructures.


In one experiment we added to the live disk embedded in a static potential for
the halo 1,000 substructures with a mass of 10$^{7}$ M$_{\odot}$ hitting the
disk from random orbits (labeled as second experiment) and found that the disk did not develop
prominent features in this situation, as expected from the analysis
of \citet{Hopk08}.
In the other experiment the same galaxy model was
evolved but with the introduction of $\approx$ 100 live dark satellites
isotropically distributed (named third experiment). 

Fig.5 illustrates the spatial distribution of the $\approx$ 100 live dark subhalos
that are placed on orbits with eccentricities ranging from 0.3 to 0.9, around the stellar disk, which is run in isolation
for 500 Myrs to achieve equilibrium prior to interacting with the substructures. The disk is not displayed in the image.

On the contrary, in the simulation where the disk is bombarded by the $\approx 100$
satellites, there is a considerable vertical random energy of the disk.
Fig.6 displays the time evolution of the first moment of the $z$
component of stars as a function of radius, $z_0$, and
$\sigma^2_z(t)-\sigma^2_z(t=0)$, the change of the vertical heating
from the initial time for the experiment with several satellites
orbiting around the disk. This result confirms the previous findings
that disk heating is caused by a few massive satellites passing
through the disk \citep[][]{Hopk08,Just15}, although also their velocity plays
a role. We note that in this case
the vertical heating is not sudden but gradually increases as time progresses. Although after 2
Gyrs there is considerable heating produced at the solar neighborhood,
the heating and flaring of stars are very pronounced in the outer part
of the disk, in agreement with the results of the disk run AQ-F2
presented in \citet{Just15}.

In this case it is possible that the several impacts of massive satellites
with different relative velocities with respect to the disk represent perturbations
with a  spectrum of higher frequencies in the range that excite 
the formation of long-bending waves that, once damped can heat the
disk and generate flares. 

\section{SUMMARY AND CONCLUDING REMARKS}

The goal of our study has been to gain physical insight into the
gravitational response of a disk to perturbations induced by
satellites passing through it, as predicted by cosmological
simulations.  We have chosen to focus mainly on the impact of one
satellite as massive as the Sagittarius dwarf galaxy in order to
isolate the phenomenon from the cumulative effects of many encounters
with smaller substructures.

We show that a not too fast satellite as massive as Sagittarius passing
through the disk can induce the disk to wobble and form patterns similar
to in-plane rings, in agreement with previous results. However, our
findings indicate that the underdense regions between rings overlap
with local enhancements in disk thickness. This coincidence is due to
stars located in underdense regions that have coupled horizontal and
vertical motions. Those stars display vertical displacements that can
extend up to 1 kpc, with non-zero vertical streaming motions as large
as 15-20 km s$^{-1}$, consistent with recent observations in the solar
neighborhood. This phenomenon appears like a local flaring.  
However, the dark halo is spherical in these numerical experiments. This assumption
might artificially slow down the precession rates of the oscillations, 
delaying the disk heating.

In the simulation with $\approx$100 satellites injected around the stellar disk 
with mass and properties expected in cosmological simulations, our results 
indicated that disk flaring is produced  in the
outer parts of the disk with significant
vertical heating experienced by the stellar disk within 2.5 Gyrs. 

Given the fact that Sagittarius likely passed through the disk and that the
Sun is located in an under-dense region, our picture naturally predicts that
in the solar neighborhood the mass density of the disk estimated by $V_z$ dispersion measurements of
stars will be over-estimated as compared to other regions of the
disk. This over-estimate, known as the problem of the Oort limit of
the stellar disk, once reconciled, might indicate a lower
mass-to-light ratio in the disk, suggesting that there is almost no need for
dark matter at the position of the Sun.

\acknowledgments 
This work is funded by NSF Grant No. AST-1211258 and ATP-NASA Grant No. NNX14AP53G.  
E.D. gratefully acknowledges support of the Alfred P. Sloan Foundation and the 
hospitality of the Aspen Center for Physics, funded by NSF Grant No. PHY-1066293.
P.M. acknowledges support by the NSF through grant AST-1229745 and NASA through grant NNX12AF87G.
L.H. acknowledges support from NASA grant NNX12AC67G and NSF grant AST-1312095.


\end{document}